\documentclass[%
 reprint,
 amsmath,amssymb,
 aps,pra
]{revtex4-2}

\usepackage[table]{xcolor}

\usepackage{siunitx}
\usepackage{graphicx}
\usepackage{dcolumn}
\usepackage{bm}
\usepackage[unicode=true, colorlinks=true, citecolor={blue!80!black}, urlcolor={blue!50!black}, linkcolor = {blue!80!black}]{hyperref}
\usepackage{easyReview} 

\usepackage{adjustbox}

\begin{document}

\preprint{APS/123-QED}

\title{Gate-tunable kinetic inductance in proximitized nanowires}

\author{Lukas Johannes Splitthoff$^{1,2}$}
\email{L.J.Splitthoff@tudelft.nl}
\author{Arno Bargerbos$^{1,2}$}
\author{Lukas Gr\"unhaupt$^{1,2}$}
\author{Marta Pita-Vidal$^{1,2}$}
\author{Jaap Joachim Wesdorp$^{1,2}$}
\author{Yu Liu$^{3}$}
\author{Angela Kou$^{4}$}
\author{Christian Kraglund Andersen$^{1,2}$}
\author{Bernard van Heck$^{5}$}

\affiliation{$^1$QuTech, Delft University of Technology, Delft 2628 CJ, The Netherlands}
\affiliation{$^2$Kavli Institute for Nanoscience, Delft University of Technology, Delft 2628 CJ, The Netherlands}
\affiliation{$^3$Center for Quantum Devices, Niels Bohr Institute, University of Copenhagen, Copenhagen 2100, Denmark}
\affiliation{$^4$Department of Physics and Frederick Seitz Materials Research Laboratory, University of Illinois Urbana-Champaign, Urbana, IL 61801, USA}
\affiliation{$^5$Leiden Institute of Physics, Leiden University, 2333 CA Leiden, The Netherlands}

\date{\today}

\begin{abstract}
We report the detection of a gate-tunable kinetic inductance in a hybrid InAs/Al nanowire. For this purpose, we have embedded the nanowire into a quarter-wave coplanar waveguide resonator and measured the resonance frequency of the circuit. We find that the resonance frequency can be changed via the gate voltage that controls the electron density of the proximitized semiconductor and thus the nanowire inductance. Applying Mattis-Bardeen theory, we extract the gate dependence of the normal state conductivity of the nanowire, as well as its superconducting gap. Our measurements complement existing characterization methods for hybrid nanowires and provide a new and useful tool for gate-controlled superconducting electronics.	
\end{abstract}

\maketitle

\section{Introduction}
Superconducting wires with high kinetic inductance \cite{Meservey1969, Annunziata2010, Zmuidzinas2012, Samkharadze2016, Niepce2019, Grunhaupt2019, Kamenov2020, Mironov2021} have found important applications as radiation detectors \cite{Mazin2005} and, more recently, in the field of quantum technology \cite{Blais2020, Kjaergaard2020, Blais2021}, where they are particularly beneficial for protected superconducting qubit designs \cite{Lin2018, Nguyen2019, Hazard2019, Gyenis2021, Gyenis2021-2, Zhang2021}.

Typically, the kinetic inductance of a nanofabricated sample is determined by the material composition and by the device geometry and, thus, is not tunable in-situ. 
Yet, some applications rely critically on the ability to tune the inductor of a quantum circuit after fabrication, e.g. for superconducting amplifiers \cite{Yamamoto2008, Krantz2019, Parker2021}. Such tunability has commonly been achieved through the use of flux-biased dc-SQUIDs \cite{Palacios-Laloy2008, Naaman2016} or current-biased conductors \cite{ Yamamoto2008, Annunziata2010, Vissers2015, Xu2019, Mahashabde2020, Parker2021}. 

A more recent source of tunability comes from the advent of hybrid semiconducting-superconducting nanostructures, which allow the realization of gate-tunable Josephson junctions \cite{deLange2015, Larsen2015, Casparis2019}. Even a semiconducting nanowire proximitized by a continuous superconducting film, without any Josephson junction, can be controlled electrostatically. Nearby gates influence the electron density in the semiconductor as well as the induced superconducting gap \cite{deMoor2018, Antipov2018, Mikkelsen2018, Winkler2019}, the two crucial parameters which control the inductive response. This electrostatic control over the properties of the proximitized nanowire suggests the possibility to realize a gate-tunable kinetic inductance by exploiting the proximitized transport channels. Such a circuit element could allow for a new class of frequency controllable resonators, amplifiers, qubits and detectors.

Here, we experimentally demonstrate a gate-tunable kinetic inductance using an InAs/Al nanowire shunting a NbTiN coplanar waveguide resonator. This system offers an easy-to-fabricate and magnetic field-compatible \cite{Kroll2019, Kringhoj2021} circuit element for superconducting electronics.
Furthermore, facilitated by the absence of etched Al segments in the nanowire, we are able to extract bulk transport properties of the hybrid InAs/Al system, which has recently attracted lots of attention for its potential use in topologically protected qubits \cite{Lutchyn2018}.

\section{Experimental setup}
\begin{figure}
	\includegraphics{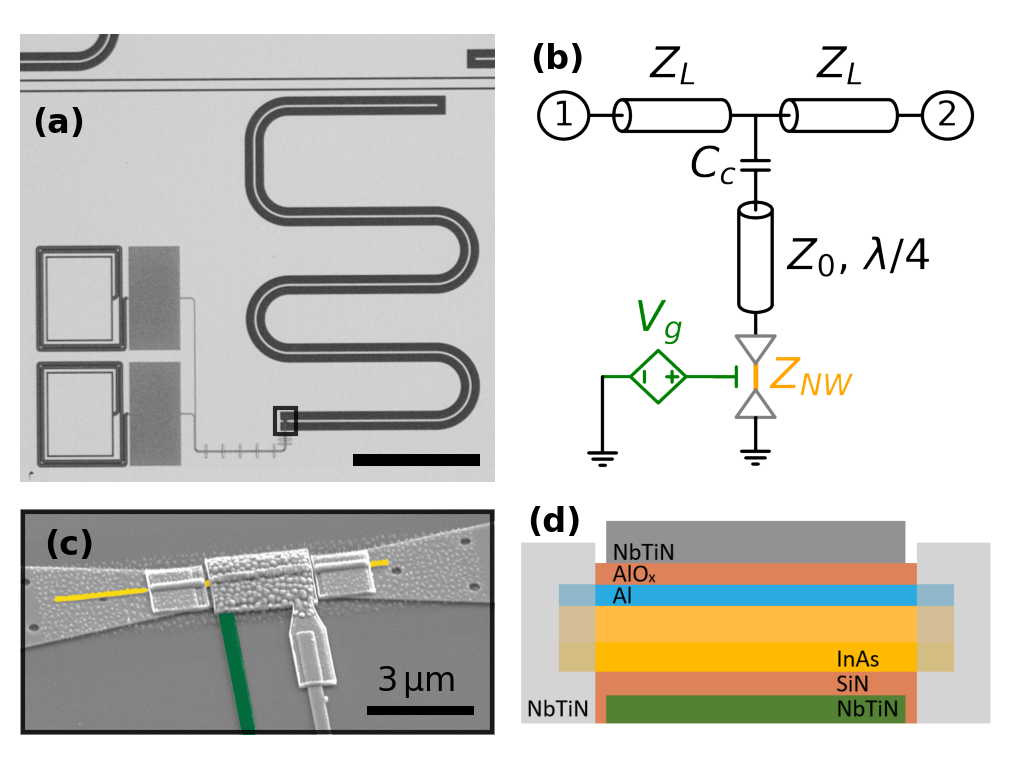} 
	\caption{Measured device and equivalent circuit. (a) Optical image of one quarter-wave coplanar waveguide resonator capacitively coupled to a feedline and shorted to ground via a proximitized nanowire. Two LC filtered pads left to the resonator connect to the gates in the vicinity of the nanowire. (scale bar \SI{300}{\micro \meter}) 
	(b) Circuit diagram of a measured device. The transmission line resonator with impedance $Z_0$ is shunted to ground by a hybrid nanowire with impedance $Z_\textrm{NW}$, which is voltage gate controlled ($V_\textrm{g}$), and capacitively coupled ($C_\textrm{C}$) to the feedline with characteristic impedance $Z_L=\SI{50}{\ohm}$. Port 1 and 2 connect to the external measurement setup.	
	(c) False colored micrograph of a proximitized nanowire (orange) which is galvanically connected to the central conductor of the resonator [black box in (a)]. The \SI{3}{\micro \meter} long nanowire section is encapsulated by bottom (green) and top (grey) gates. For this experiment we only use the bottom gate.  
	(d) Schematic longitudinal cut away of InAs nanowire (yellow) with continuous Al shell (blue) between two NbTiN contacts. Bottom (green) and top (grey) gates with dielectric (orange) define the electrostatic potential at the nanowire.}
	\label{fig: nanowire resonator}
\end{figure}

We employ a quarter-wave coplanar waveguide resonator which is capacitively coupled to a feedline and shorted to ground by the proximitized nanowire, see Fig. \ref{fig: nanowire resonator}a. 
The $l=\SI{3}{\micro \meter}$ long nanowire section is galvanically connected to the central conductor of the resonator and to ground, see Fig. \ref{fig: nanowire resonator}c. The nanowire is encapsulated by bottom and top gates for electrostatic control. In this experiment we use the bottom gate only, which extends along the entire nanowire section (optically hidden by top gate), as illustrated in Fig. \ref{fig: nanowire resonator}d. Each gate line is filtered by on-chip LC filters \cite{Mi2017}.
The schematic longitudinal cross section in Fig. \ref{fig: nanowire resonator}d highlights the continuous Al shell on two facets of the gated InAs nanowire, which is connected to the NbTiN circuit (App. \ref{app: fabrication}). 

To characterize the bare resonator properties, we use an identical reference resonator, in which the nanowire is replaced by a continuous \SI{150}{\nano \meter} thick NbTiN film (App. \ref{app: reference resonator}).
Multiple nanowire and reference resonators are frequency multiplexed on the same chip and measured at \SI{15}{\milli \kelvin} base temperature inside a dilution refrigerator. In the main , we focus on one nanowire (two-facet Al-InAs nanowire, \SI{110(5)}{\nano \meter} diameter, \SI{6}{\nano \meter} thick Al shell) and one reference resonator which exemplify trends observed in 12 different devices (App. \ref{app: Nanowire resonators}).

\section{Gate-tunable kinetic inductance} 
\begin{figure}
	\includegraphics{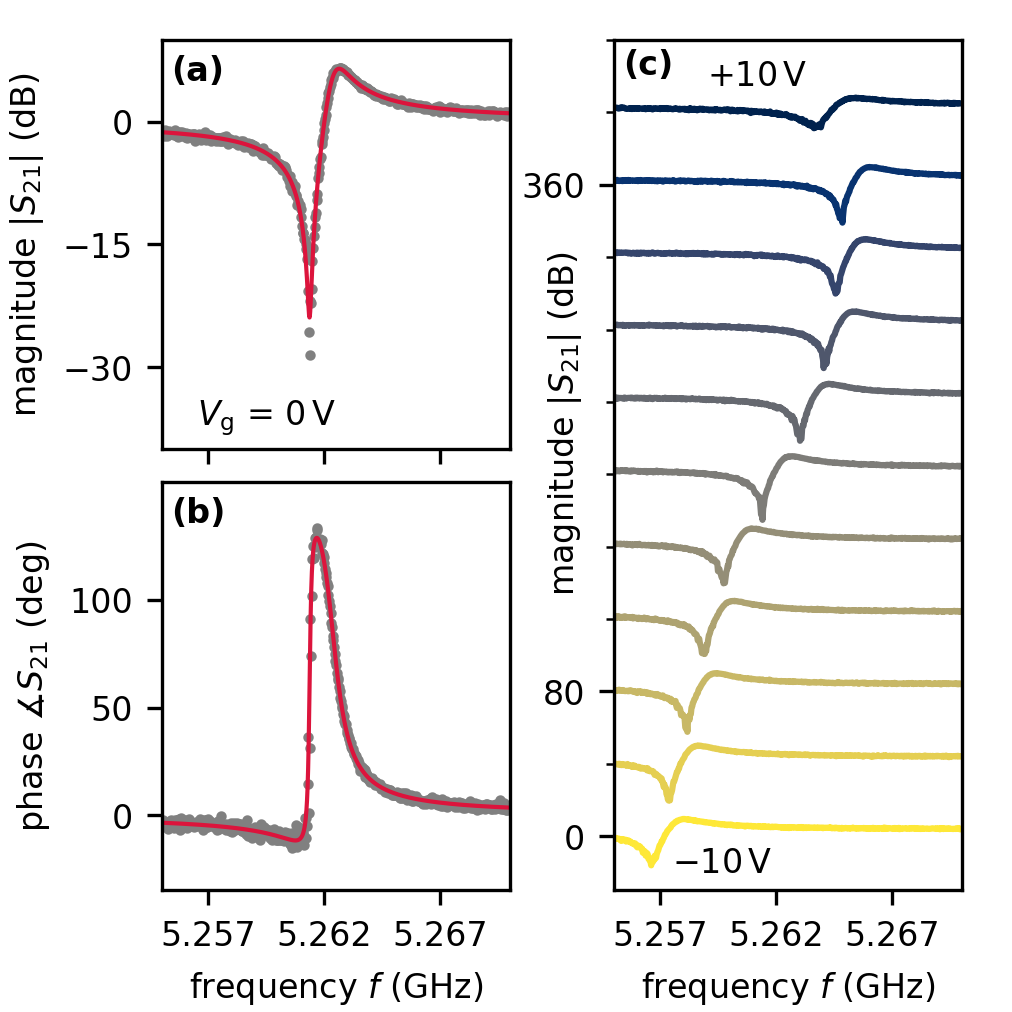}
	\caption{Gate dependent frequency shift of the nanowire resonator. (a,b) Typical magnitude and phase of transmission parameter $S_{21}$, respectively, of a single nanowire resonator for a fixed gate voltage $V_\textrm{0}=\SI{0}{\volt}$. Red line shows fit result. (c) Magnitude of the transmission $S_{21}$ for 11 different, linear spaced gate voltages in the range from \SI{-10}{\volt} to \SI{10}{\volt}. Each line with a fixed gate voltage is offset by \SI{40}{dB} for better visibility. }
	\label{fig: raw data}
\end{figure}

The nanowire resonator system can be described by the circuit model in Fig. \ref{fig: nanowire resonator}b: a transmission line with characteristic impedance $Z_0$ is shunted by a gate-voltage-dependent ($V_\textrm{g}$) impedance $Z_\textrm{NW}$, and capacitively coupled ($C_\textrm{C}$) to a common feedline ($Z_\textrm{L}=\SI{50}{\ohm}$). We measure the transmission parameter $S_{21}$ in the vicinity of the resonance frequency. The amplitude and phase of $S_{21}$ display typical resonant behaviour, as shown in Fig. \ref{fig: raw data}a,b. We use a linear resonator model \cite{Khalil2012,Probst2015} to fit the resonance frequency $f_\textrm{r}$ and internal quality factor $Q_\textrm{i}$. 

\begin{figure*}[t]
	\includegraphics{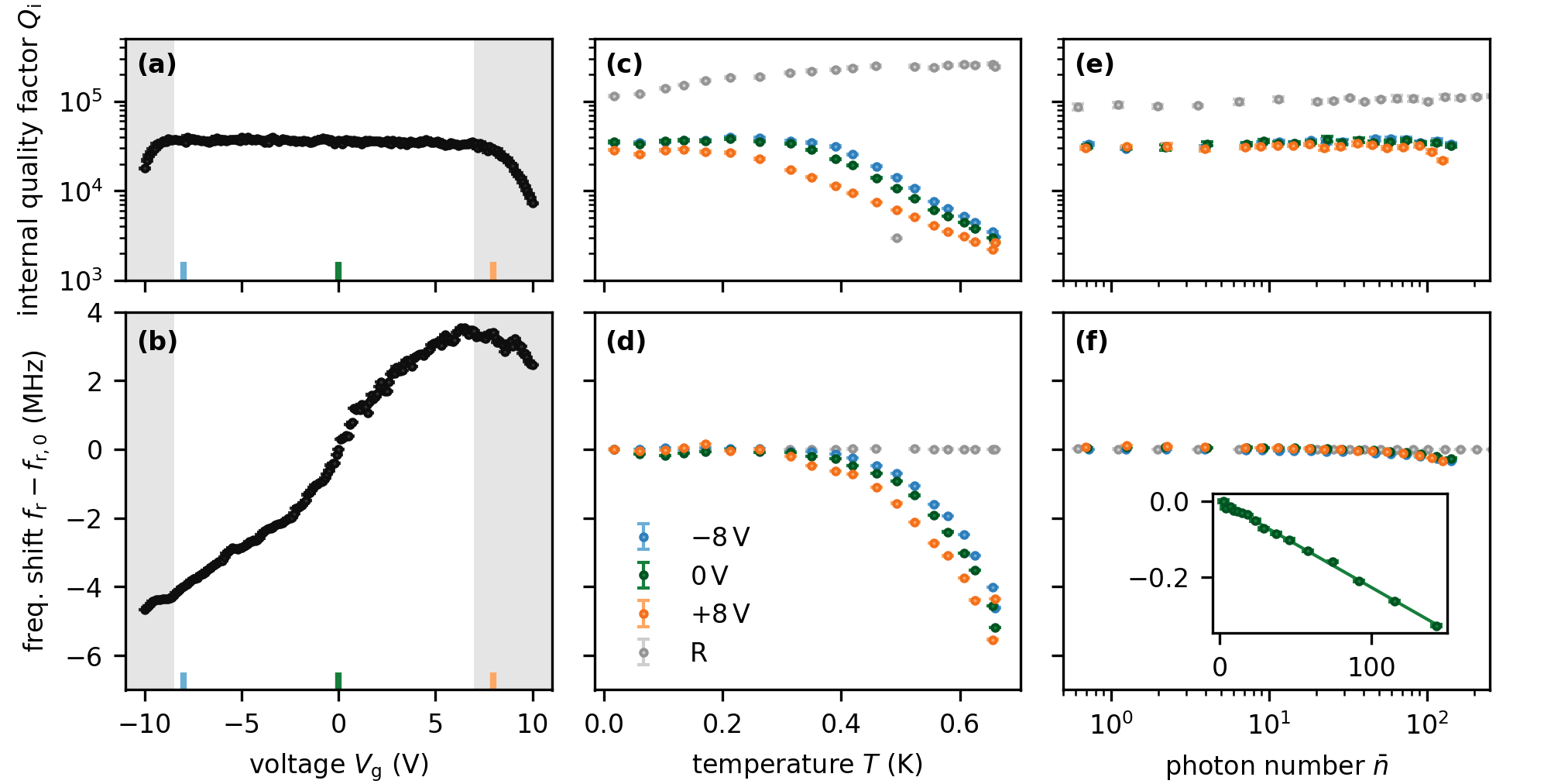}
	\caption{Resonator response versus control parameters (a,b) Internal quality factor and frequency shift as function of gate voltages extracted from fits to $S_{21}$ data (cf. Fig. \ref{fig: raw data}a,b). The grey boxes indicate the regions of decreasing $Q_\textrm{i}$ for $V_\textrm{g}<\SI{-8.5}{\volt}$ and $V_\textrm{g}>\SI{7}{\volt}$ (see main text). Coloured lines at the bottom mark gate points used in panels (c-f). (c,d) Internal quality factor and frequency shift for the nanowire resonator versus temperature for different gate voltages relative to their zero temperature values. Reference resonator (R) data shown for comparison. (e,f) Internal quality factor and frequency shift for the nanowire resonator versus average intra-cavity photon number for different gate voltages. For higher photon number the nanowire resonator becomes increasingly non-linear and the linear fit model does not capture the measured data above $\bar{n}\gtrsim 200$. Reference resonator (R) data shown for comparison. The error bars indicate the $1\sigma$ confidence interval of fit. Inset: Zoom-in on frequency shift of nanowire resonator at \SI{0}{\volt} with linear fit.}
	\label{fig: extracted resonator parameters}
\end{figure*}

We observe a monotonic increase in resonance frequency as a function of gate voltage up to \SI{8}{\volt} (Fig. \ref{fig: raw data}c). The shift is about \SI{8}{\mega \hertz} over the total gate range available for measurement (\SI{-10}{\volt} to \SI{10}{\volt}). In Fig. \ref{fig: extracted resonator parameters}a,b, we show the extraction of $f_\textrm{r}$ and $Q_\textrm{i}$ as a function of gate voltage. 
The internal quality factor is nearly constant in the entire range, but it decreases sharply at large applied voltages, an effect which we attribute to the breakdown of the gate dielectric (App. \ref{app: gate leakage}).

When the temperature is increased, the resonance frequency and internal quality factor decrease, at all gate voltages, see Fig. \ref{fig: extracted resonator parameters}c,d. We attribute this behaviour to the suppression of superconductivity in Al. 
Between \SI{20}{mK} and \SI{750}{mK}, the frequency is reduced by about \SI{6}{MHz}, while $Q_\textrm{i}$ drops by one order of magnitude.
We notice that for both quantities the temperature response exhibits a gate dependence: it is stronger for positive gate voltages, which, as shown below, we attribute to a change in the induced superconducting gap in InAs.
In contrast, in the same temperature range, the resonance frequency of the reference resonator remains approximately constant, while $Q_\textrm{i}$ increases slightly. These observations are consistent with the much higher critical temperature of NbTiN and the saturation of two-level systems contributing to dielectric losses \cite{Bruno2015}.

The resonance frequency decreases linearly with the number of intra-resonator photons, $\bar{n}$, as expected in the presence of a Kerr non-linearity arising from the inductance \cite{Tholen2009-2, Grunhaupt2018}. 
The frequency shift relative to the single photon level (App. \ref{app: setup}) is about \SI{0.2}{\mega \hertz} at \SI{-120}{dBm} input power, corresponding to $\bar{n}\approx \SI{100}{}$ (Fig. \ref{fig: extracted resonator parameters}e,f), one order of magnitude smaller than the frequency shifts observed versus gate voltage or temperature. 
We measure a Kerr coefficient $K/2\pi=(f_\textrm{r}-f_\textrm{r,0})/\bar{n}=\SI{2.28(3)}{\kilo \hertz}$ (inset Fig. \ref{fig: extracted resonator parameters}f), which is sufficient for the realization of superconducting parametric amplifiers  \cite{Tholen2009-2, Tholen2009}.
For larger photon numbers, higher order non-linear processes set in and the resonator bifurcates \cite{Vijay2009,Andersen2020}, thus the linear model does not capture the measured data for $\bar{n}\gtrsim 200$ (App. \ref{app: Resonator bifurcation}).
The reference resonator inductance is not susceptible to these photon numbers, thus there is no frequency shift. $Q_\textrm{i}$ increases with microwave power which is consistent with the saturation of two-level systems \cite{Bruno2015}. 
We thus choose 100 intra-resonator photons for the remaining measurements to have sufficient signal-to-noise ratio without inducing strong nonlinear effects on the nanowire.

\section{Bulk properties of the proximitized nanowire}
\begin{figure}
	\includegraphics{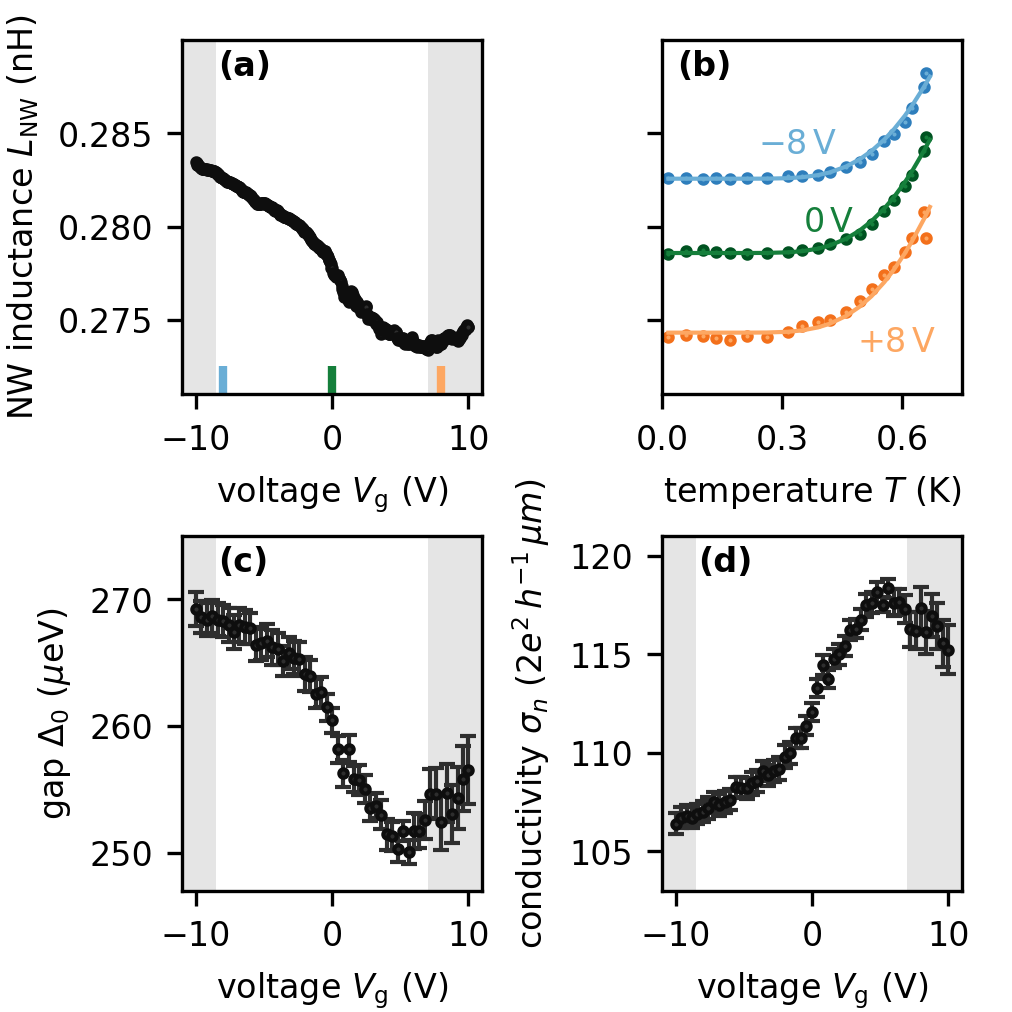}
	\caption{Extracted nanowire parameters from resonator data shown in Fig \ref{fig: extracted resonator parameters}. (a) Inductance of the nanowire resonator versus gate voltage extracted from loaded resonator model (see main text). (b) Nanowire resonator inductance versus temperature for three different gate voltages. Lines indicate fit to Matthis-Bardeen model. (c) Superconducting gap $\Delta_0$ extracted from the fit to the temperature dependence for each gate voltage. (d) Normal state conductivity $\sigma_n$ extracted from the fit to the temperature dependence for each gate voltage. The greyed-out areas indicate the regions of decreasing $Q_\textrm{i}$.}
	\label{fig: extracted nanowire parameters}
\end{figure}

We use the measured frequency shifts to extract relevant physical properties of the hybrid nanowire. Assuming that the imaginary part of the nanowire impedance is purely inductive, $\Im (Z_\textrm{NW}) = 2 \pi f_\textrm{NW} L_\textrm{NW}$, which is justified given the electromagnetic field profile along the resonator in the limit $Z_\textrm{NW}\ll Z_0$, a loaded transmission line resonator model (App. \ref{app: nanowire impedance}) yields the nanowire inductance $L_\textrm{NW}=Z_0/4 (f_\textrm{NW}^{-1}-f_{0}^{-1})$, where $f_\textrm{NW}$ is the resonance frequency of the loaded nanowire resonator and $Z_0$ and $f_{0}$ are the characteristic impedance and resonance frequency of the unloaded coplanar waveguide resonator.
Using a conformal mapping technique \cite{Simons2001} and the average kinetic sheet inductance of the NbTiN film obtained from the measurement of four reference resonators, we determine $Z_0=\SI{134.2(4)}{\ohm}$ and $f_{0}=\SI{5.502(16)}{GHz}$. The uncertainties of these quantities mainly depend on the uncertainty of the NbTiN kinetic sheet inductance. Thus, we can extract $L_\textrm{NW}$ from the measured $f_\textrm{NW}$, as shown in Fig. \ref{fig: extracted nanowire parameters}a.

At zero voltage, the nanowire inductance is $L_\textrm{NW}=\SI{278(17)}{\pico \henry}$ with about \SI{6}{\percent} constant systematic error stemming from the uncertainties on $Z_0$ and $f_0$. $L_\textrm{NW}$ varies by $\SI{3.5}{\percent}$ over the range from \SI{-10}{\volt} to \SI{7}{\volt} before superconductivity is suppressed by gate leakage. 
The geometric inductance associated with the nanowire segment can be estimated to be $L_\textrm{NW,geo}=\SI{4.5}{\pico \henry}$ (App. \ref{app: geometric inductance}), thus we conclude that the shunt impedance is dominated by the kinetic inductance of the hybrid nanowire.

Using the same method, we extract the temperature dependence of $L_\textrm{NW}$, as shown in Fig. \ref{fig: extracted nanowire parameters}b. For all gate voltages, the nanowire inductance increases with temperature. According to the Mattis-Bardeen theory \cite{Mattis1958, Tinkham2004} in the low frequency limit, $hf \ll \Delta_0$, the kinetic inductance is given by
\begin{align} \label{eq: MB thinfilm inductance}
	L_\textrm{NW}=\frac{\hbar l}{\pi \sigma_\textrm{n} \Delta_0}\tanh^{-1}\bigg(\frac{\Delta_0}{2 k_\textrm{B} T}\bigg),
\end{align} 
where $\Delta_0$ is the superconducting gap and $\sigma_\textrm{n}$ is the one-dimensional normal state conductivity of the hybrid nanowire. Note that the latter absorbs the cross-sectional electron density distribution and has physical dimensions of conductance times length. The increase of $L_\textrm{NW}$ with temperature can be used to extract $\Delta_0$ as a function of gate voltage.

Increasing the gate voltage, the extracted superconducting gap decreases by \SI{7}{\percent} from \SI{270(1)}{\micro\electronvolt} to \SI{251(1)}{\micro\electronvolt} (Fig. \ref{fig: extracted nanowire parameters}c). We note that at higher temperature, the onset of gate leakage shifts down to \SI{5}{\volt} causing an increase in the fit uncertainty and interrupting the monotonous decrease of the gap with gate voltage.
Moreover, we observe two distinct slopes: a shallow slope between \SI{-10}{\volt} and \SI{-2}{\volt} and a steeper slope between \SI{-2}{\volt} and \SI{5}{\volt}. Similar observations have also been made in previous experiments on similar devices \cite{Antipov2018,Mikkelsen2018,deMoor2018,Winkler2019,Shen2021}. 
We attribute the first regime to the strong hybridization of the InAs sub-bands with Al, and the second to an increase in the density in InAs with the occupation of new sub-bands. We do not observe a plateau in the bulk properties at negative gate voltages, which indicates that we have not reached the full depletion of the semiconductor. Whether the latter is possible at all depends on the value of the band-offset pinning the conduction band of InAs at the InAs/Al interface. If the band offset is too large, it may not be possible to reach full depletion by decreasing the gate voltage further, since eventually the gate may be screened by the accumulation of holes on the facets of the InAs nanowire closer to the gate electrode. \cite{Antipov2018}.
We note that even over this large gate voltage range, the extracted gap remains finite. This observation does not rule out the possible onset, at positive gate voltages, of unproximitized states on the opposite side of Al-covered nanowire facets \cite{Winkler2019}. 
These states, if present, would not contribute considerably to the kinetic inductance.

Having determined the gate dependence of $\Delta_0$, we can use Eq. (\ref{eq: MB thinfilm inductance}) to extract $\sigma_\textrm{n}$ using the values of $L_\textrm{NW}$ measured at the lowest temperature (Fig. \ref{fig: extracted nanowire parameters}d). 
The resulting normal state conductivity increases by \SI{10}{\percent} over the gate voltage range. This change in conductivity translates, for a nanowire length of \SI{3}{\micro \meter}, to an increase in conductance of $\approx\SI{3.9}{\, 2e^2\, h^{-1}}$, giving a lower bound of 4 proximitized conduction channels added under the Al shell.
Our measurement also yields a normal state resistance at \SI{-10}{\volt} of $R_\textrm{n}=\SI{365(1)}{\ohm}$ or $R_\Box=\SI{13.4(1)}{\ohm \, \Box^{-1}}$, which is consistent with the expected values for a \SI{3}{\micro \meter} long and \SI{110}{\nano \meter} wide Al thin film \cite{Phan2021}.
The increase of conductivity with gate voltage overcompensates the simultaneous decrease of $\Delta_0$ to determine the decrease of the inductance $L_\textrm{NW}$ with gate voltage.

We note that the Mattis-Bardeen theory is formulated for a simple BCS superconductor and thus it does not capture the fact that the induced gap in the semiconductor could be different from that in the Al shell \cite{Kiendl2019}, and possibly varies among the proximitized sub-bands \cite{Reeg2017, Reeg2018, Antipov2018, Mikkelsen2018, Woods2018}. While these differences may cause strong deviations from the Mattis-Bardeen theory regarding the microwave absorption spectrum at frequencies comparable to the gap, the low-frequency inductive response of the condensate is likely to be less sensitive to the precise profile of the density of states. As a straightforward ansatz, it is therefore sensible to test the applicability of Eq. (\ref{eq: MB thinfilm inductance}) to the data. A posteriori, the results reveal gate voltage trends that are consistent with those expected from detailed microscopic modeling of the field effect in the InAs/Al system \cite{Woods2018, Kiendl2019, Menard2019}, thus validating this simple approach. However, in order to fit the gate dependence of the data to microscopic parameters such as the hybridization strength between the two materials, an extension of the Mattis-Bardeen theory to proximitized systems would be required.

\section{Conclusions}
This experiment complements existing characterization methods for hybrid nanowires, which have been primarily based on low-frequency transport techniques \cite{Mourik2012,Gul2015,vanWeperen2015,Albrecht2016}.
The measurement of the microwave kinetic inductance allows us to directly probe the bulk properties of the hybrid nanowires, in contrast with measurement on devices with etched Josephson junctions, which are affected by the junction geometry and fabrication details. Given that the kinetic inductance gives access to the normal state conductivity in the proximitized nanowire, our method, if complemented with a capacitive measurement of the electron density via the bottom gate \cite{Malinowski2021}, could allow to extract the mobility of the semiconductor under the Al shell. Such a measurement could be of importance to determine whether the hybrid system meets the stringent disorder requirements for Majorana applications \cite{Ahn2021}.
The value of resonator-based material characterisation in the context of hybrid systems has recently been demonstrated in 2DEGs \cite{Phan2021}. 

Besides the fundamental interest to characterize hybrid nanowires, the presented approach offers a promising path towards gate-controlled superconducting electronics. In particular, the low loss and finite Kerr non-linearity will allow for frequency tunable resonators, switches, microwave detectors and parametric amplifiers \cite{Wustmann2013}. In view of the potential application as frequency tunable resonators we report the frequency stability and hysteresis in the Appendix \ref{app: stability}. 
The tunable frequency range can be enhanced by optimizing the choice of materials (e.g. including dielectrics with larger dielectric constants),
the design of the gate electrodes, and the participation ratio of the nanowire inductance to the total resonator inductance.
\\The raw data and the data analysis code at the basis of the results presented in this work are available online \cite{Splitthoff2022}.

\section*{Acknowledgements} 
We thank Peter Krogstrup for the supervision of the materials growth and Leo P. Kouwenhoven for support and discussion.
We also thank Andrey Antipov and Andrew Higginbotham for their careful feedback on the manuscript.
This research was co-funded by the allowance for Top consortia for Knowledge and Innovation (TKI) from the Dutch Ministry of Economic Affairs and the Microsoft Quantum initiative. 
CKA and BvH acknowledge financial support from the Dutch Research Council (NWO).

\section*{Author contributions}
LJS, BvH and AK conceived the experiment. LJS fabricated the devices with help from MPV. Nanowires were grown by YL. LJS acquired and analysed the data with input from AB, LG, AK, CKA and BvH. LJS, CKA and BvH wrote the manuscript with input from all other co-authors. AK, CKA and BvH supervised the project.

\appendix

\section{Fabrication} \label{app: fabrication}
We fabricate the resonator circuit and the gate lines from a \SI{35}{\nano \meter}-thick sputtered NbTiN film on low pressure chemical vapor deposition (LPCVD) SiN on Si. The kinetic sheet inductance of \SI{5.31(5)}{\pico \henry \, \Box^{-1}} has been measured using four reference resonators multiplexed to the same feedline, see Sec. \ref{app: reference resonator}. We pattern the ground plane using e-beam lithography and SF$_6$/O$_2$ based reactive ion etching. \SI{30}{\nano \meter}-thick plasma enhanced chemical vapour deposition (PECVD) SiN serves as bottom gate dielectric. We transfer the two-facet InAs/Al nanowire on top of the SiN bottom gate with a nano-manipulator. The InAs nanowires were grown by vapor–liquid–solid (VLS) growth with a diameter of \SI{110(5)}{\nano \meter}, and nominal thickness of the Al of \SI{6}{\nano \meter} \cite{Krogstrup2015}. We electrically contact the nanowires to the circuit via lift-off defined \SI{150}{\nano \meter}-thick sputtered NbTiN leads after Ar milling which reduces the contact resistance. Similarly, we add top gates made from AlOx dielectrics and NbTiN to the design to overcome a possible lack of tunability due to the uncontrolled alignment of the Al shell with respect to the gates, but do not use them in this experiment as they got shorted to ground in a subsequent fabrication step. The grains visible in the SEM of the nanowire segment in the main text originate from polymer resist residuals.

\section{Measurement setup} \label{app: setup}
\begin{figure}
	\includegraphics[scale=0.5]{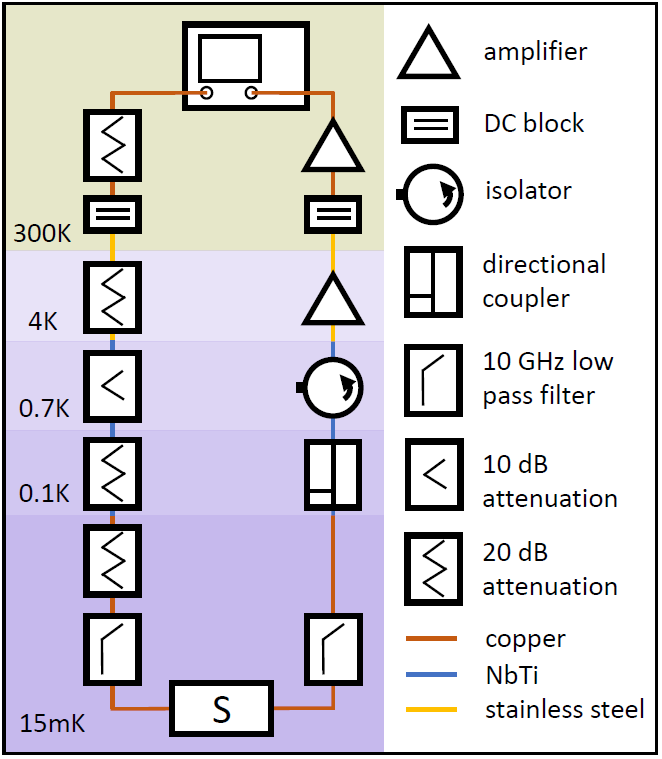}
	\caption{Transmission measurement setup. A vector network analyser connected to highly attenuated input and amplified output line probes the sample (S) with resonators.}
	\label{fig: microwave setup}
\end{figure}

For the microwave spectroscopy, we use a RF transmission measurement setup installed in a dilution refrigerator operating at \SI{15}{\milli \K} base temperature, see Fig. \ref{fig: microwave setup}. We probe the sample with resonators using a vector network analyser (VNA). 
The input line consists of microwave attenuators at room temperature (\SI{-20}{dB}) and at cryogenic temperatures (\SI{-70}{dB}). In addition, at each of the four temperature stages eccosorb filters with about \SI{0.85}{dB} attenuation per \SI{10}{GHz} are installed. The cables connecting the VNA to the \SI{4}{\K} plate add about \SI{8}{dB} of attenuation. Then, NbTi cables connect to the \SI{15}{\milli \K} plate. We estimate the total input attenuation up to the sample plane to be \SI{-105(5)}{dB} when accounting for the additional insertion loss of filters and other cable connections to the sample mount. The \SI{5}{dB} uncertainty arises from the summation of the individual components measured at room temperature.
Based on this attenuation we compute the intra-resonator photon number \cite{Bruno2015}
\begin{align}
	\bar{n}=\frac{2 Q_t^ 2}{2 \pi f_r Q_c}  \frac{P_{in}}{h f_r},
\end{align}
where $Q_\textrm{t}$ is the total quality factor, $Q_\textrm{c}$ is the coupling quality factor and $P_\textrm{in}$ is the power at the input port of the sample. The intra-resonator photon number deviates from the true value by not more than a factor of \SI{3.2}{} due to the uncertainty of the input attenuation. 

The 2x\SI{7}{\milli \meter} chip on which the measurements are performed is glued with silver paint onto a gold plated copper mount and electrically connected to a printed circuit board using aluminium wire-bonds. 

The output signal is amplified using a high electron mobility transistor amplifier at the \SI{4}{\K} stage. An isolator with \SI{42}{dB} typical isolation prevents reflections. The directional coupler in the output line couples to an unused microwave line. The \SI{10}{GHz} low pass filter at both input and output lines clean up the spectrum. DC blocks are placed at the breakout point of the fridge. A room temperature amplifier amplifies the signal before it enters the receiver port of the VNA.  

The electrostatic gates are controlled via DC lines using a QuTech IVVI rack. The DC lines are low pass filtered at the mixing chamber stage.

\section{Resonator fitting} \label{app: resonator fitting}
We use a background corrected linear notch type resonator model \cite{Flanigan2021} to fit the acquired resonator data. The model is based on the scattering parameter of the resonator
\begin{align}
	S_{21}=1 - \frac{Q_\textrm{c}^{-1} (1+i a)}{Q_\textrm{c}^{-1} + Q_\textrm{i}^{-1} + 2i df},
\end{align}
where $df = f/f_\textrm{r}-1$ is the frequency ratio, $a$ is the asymmetry arising from the complex loading of the resonator, $Q_\textrm{c}$ is the coupling quality factor and $Q_\textrm{i}$ is the internal quality factor  \cite{Khalil2012,Probst2015}.  
We correct the scattering parameter of the resonator by multiplying a general background model to $S_{21}$ and obtain the system scattering parameter 
\begin{align}
	\tilde{S}_{21} = \big(m_0+m_1 f\big) e^{2\pi if\tau +i \varphi} S_{21},
\end{align}
which we fit to the data. The background model includes a linear magnitude with offset $m_0$ and slope $m_1$ and a linear phase with offset $\varphi$ and slope $\tau$.

\section{Extraction of nanowire impedance} \label{app: nanowire impedance}
We employ a loaded transmission line resonator model \cite{Pozar2012} to extract the impedance of the nanowire $Z_\textrm{NW}$ from the measurement of the resonant frequency $\omega_\textrm{NW}$ and internal quality factor $Q_\textrm{i}$ of the nanowire resonator.
The input impedance of a transmission line of length $l_0$ loaded by an impedance $Z_\textrm{NW}$ can be written as
\begin{align} \label{eq: loaded transmission line resonator model}
	Z_\textrm{in} = Z_0 \,\frac{Z_\textrm{NW} + Z_0 \tanh(\alpha_0 l_0 + i \beta l_0)}{Z_0 + Z_\textrm{NW} \tanh(\alpha_0 l_0 + i \beta l_0)}\,.
\end{align}
Here, $Z_0$ is the frequency-dependent characteristic impedance of the transmission line, $\alpha_0$ is the absorption coefficient, and $\beta$ is the frequency-dependent propagation coefficient.
The use of Eq.~\eqref{eq: loaded transmission line resonator model} is justified for our devices since the length of the nanowire is much shorter than that of the resonator.

We are interested in the input impedance close to the fundamental frequency of the nanowire resonator.
In the limit of a small nanowire impedance, this frequency will be close to the fundamental frequency of the unloaded resonator, $\omega_0$, which can be measured via the reference resonators, as discussed in the main text.
We therefore expand Eq.~\eqref{eq: loaded transmission line resonator model} close to this frequency. Furthermore, we assume that the intrinsic losses are small, $\alpha_0l_0\ll 1$, thus the characteristic impedance reduces to $Z_0=\sqrt{L/C}$, where $L$ is the series inductance and $C$ is the shunt capacitance per unit length of the transmission line. In the limit $|Z_\textrm{NW}|\ll Z_0$ we obtain:
\begin{equation}
	Z_\textrm{in} \approx \frac{2\omega_0 Z_0}{\pi}\,\frac{1}{\tfrac{1}{2}\Gamma + i (\omega - \omega_\textrm{NW})}
\end{equation}
where $\omega_\textrm{NW}$ is the shifted resonant frequency of the nanowire resonator,
\begin{equation}\label{eq:omega_nw}
	\omega_\textrm{NW} = \omega_0 \,\left[1 - \frac{2}{\pi}\frac{\Im(Z_\textrm{NW})}{Z_0}\right]
\end{equation}
and $\Gamma$ is the full-width at half-maximum, the sum of a contribution due to the intrinsic losses along the transmission line and a contribution coming from the nanowire:
\begin{equation}
	\Gamma = \Gamma_0 + \Gamma_\textrm{NW} = \frac{4\omega_0}{\pi}\,\left[\alpha_0l_0 + \frac{\Re(Z_\textrm{NW})}{Z_0}\right]\,.
\end{equation}
The total internal quality factor of the nanowire resonator can therefore be given as:
\begin{equation}\label{eq:qfactor}
	Q_\textrm{i}^{-1} = Q_0^{-1} + Q_\textrm{NW}^{-1}
\end{equation}
with $Q_0 = \omega_0/\Gamma_0$ and $Q_\textrm{NW}=\omega_0/\Gamma_\textrm{NW}$.
Putting Eqs.~\eqref{eq:omega_nw} and \eqref{eq:qfactor} together, we obtain the nanowire impedance as a function of the measurable quantities $\omega_0, \omega_\textrm{NW}, Q_\textrm{i}, Q_0$:
\begin{equation}
	Z_\textrm{NW} =  \frac{\pi Z_0}{4} \left(\frac{1}{Q_\textrm{i}}-\frac{1}{Q_0}\right) + i \frac{\pi Z_0}{2} \frac{\omega_0-\omega_\textrm{NW}}{\omega_0}
\end{equation}
This number has positive real and imaginary part for $ Q_0 \ge Q_\textrm{i} $ and $ \omega_0 \ge \omega_\textrm{NW}$.

\section{Complementary measurements}

\subsection{Nanowire resonators} \label{app: Nanowire resonators}

\begin{figure*}
	\includegraphics{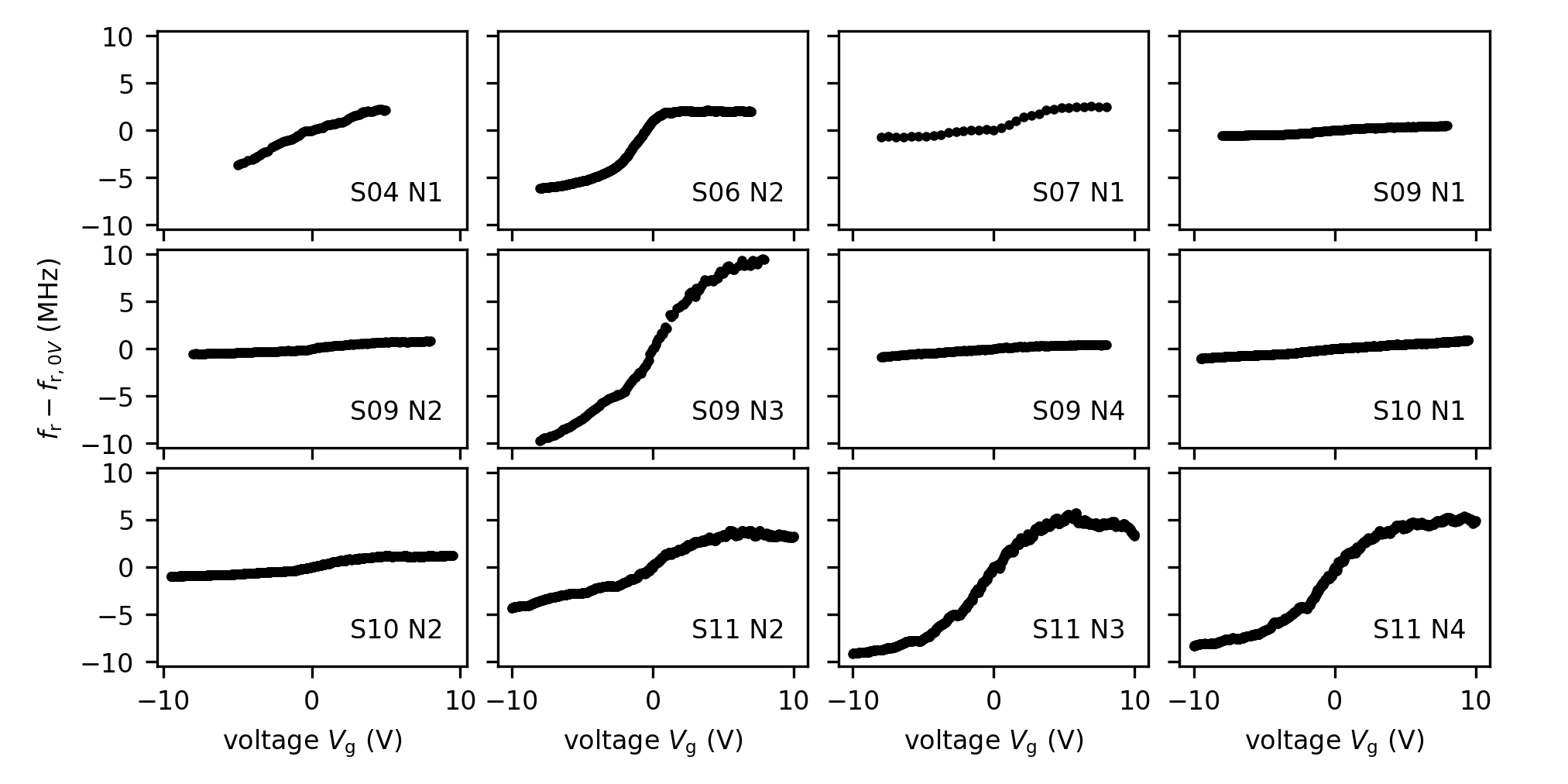}
	\caption{Frequency shift as function of gate voltages extracted from fits to $S_{21}$ data for 12 different nanowire resonators. The offset $f_\textrm{NW,0V}$ is listed in Tab. \ref{tab: nanowire resonator parameters overview}.}
	\label{fig: statistics}
	\vspace*{\floatsep}
	\includegraphics{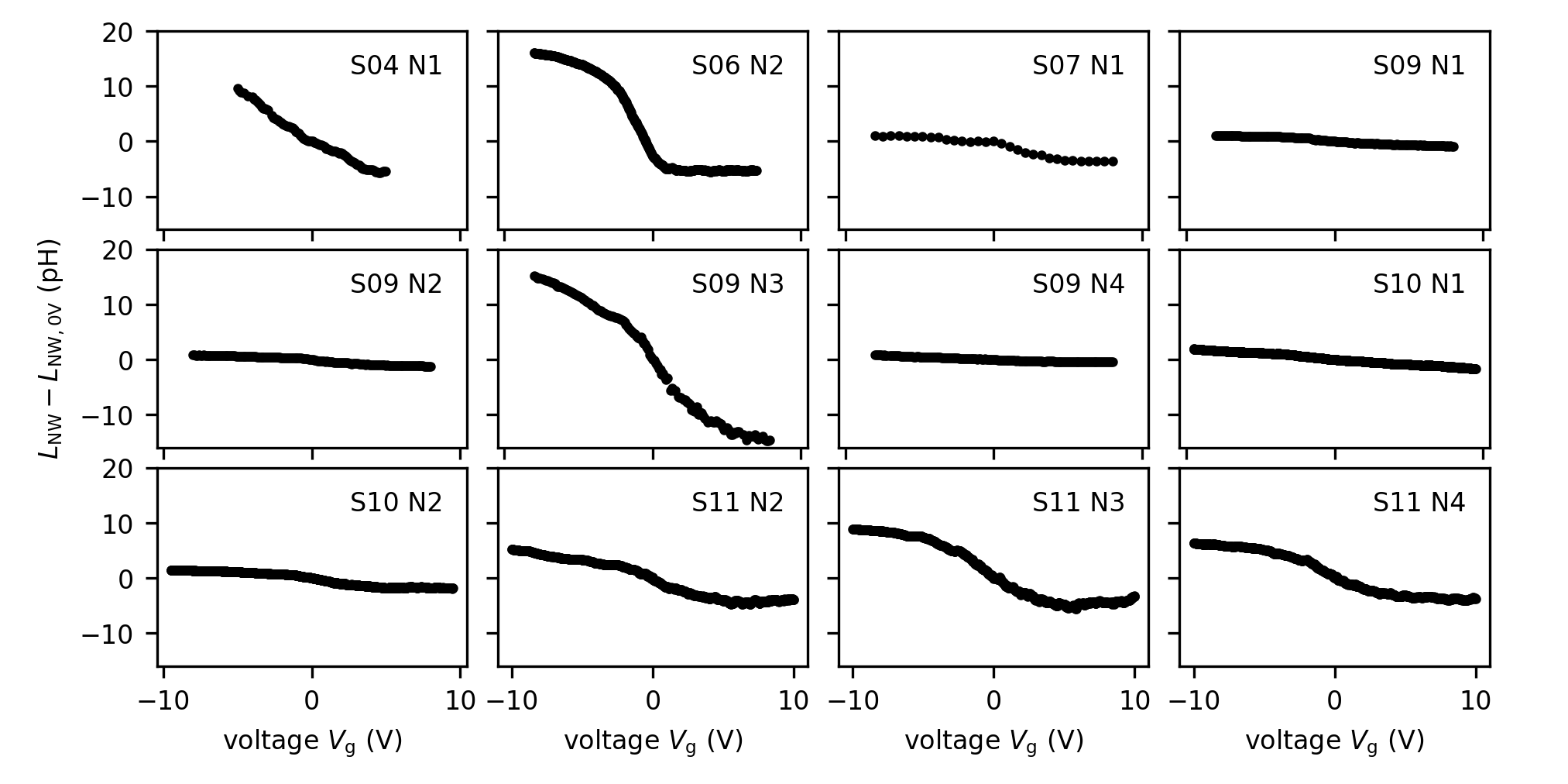}
	\caption{Change in nanowire inductance as function of gate voltages for 12 different nanowire resonators. The offset $L_\textrm{NW,0V}$ is different for each nanowire and listed in Tab. \ref{tab: nanowire resonator parameters overview}. }
	\label{fig: statistics_inductance}
\end{figure*}

During the course of this work, we prepared 11 samples with 4 nanowire resonators and 4 reference resonators each. 
In Tab. \ref{tab: nanowire resonator parameters overview} and Fig. \ref{fig: statistics} and Fig. \ref{fig: statistics_inductance} we report the parameters and basic characterization of all 12 nanowire resonators exhibiting a gate voltage dependence. Other nanowire resonators were discarded because of shorted gates, highly resistive NbTiN to nanowire contacts or broken nanowires. 
The data which entered the main text is taken on a single nanowire resonator N2 and reference resonator R1 from sample S11. We consider the data set of S11 as representative for the entire set of measurements. 

We design the nanowire resonators starting from the CPW resonators with $Z_0 \gg Z_\textrm{NW}$ by adjusting the geometry, in particular width $w$, spacing $s$ and length $l$ of the CPW transmission line. The resonator parameters are listed in Tab. \ref{tab: nanowire resonator parameters overview}. For all nanowire resonator the coupling quality factor is $Q_\textrm{c}=\SI{5.5e3}{}$. Based on the kinetic sheet inductance of NbTiN, we calculate the nanowire inductance. We also summarize further key quantities to compare the different nanowires. 
\begin{table}
	\begin{adjustbox}{width=\columnwidth,center}
		\begin{tabular}{c|c|c|c|c|c|c|c|c|c}
			\rule{0pt}{10pt} label & $w$      & $s$      & $l$        & $L_\textrm{k}$          & $L_\textrm{NW}$    & $f_\textrm{NW}$  & $\Delta V_\textrm{g}$ & $\Delta f_\textrm{NW}$  & $\Delta L_\textrm{NW}$  \\ \hline
			\rule{0pt}{10pt}       & \si{\micro \meter}  & \si{\micro \meter}   & \si{\micro \meter}     & \si{\pico \henry \, \Box^{-1}}      & \si{\nano \henry}        & \si{\giga \hertz}    & \si{\volt}    & \si{\mega \hertz} & \si{\pico \henry}\\ \hline \hline
			\rule{0pt}{10pt} S04 N1    & \SI{5}{} & \SI{20}{} & \SI{4898}{} & \SI{5.50}{} & \SI{0.45}{} & \SI{3.629}{} & \SI{10}{} & \SI{5.9}{} & \SI{15}{}\\
			
			\rule{0pt}{10pt} S06 N2    & \SI{5}{} & \SI{20}{} & \SI{4948}{} & \SI{5.67}{} & \SI{0.31}{} & \SI{3.617}{} & \SI{15}{} & \SI{8.2}{} & \SI{22}{}\\
			
			\rule{0pt}{10pt} S07 N1    & \SI{4}{} & \SI{20}{} & \SI{3200}{} & \SI{5.45}{} & \SI{0.62}{} & \SI{5.087}{} & \SI{16}{} & \SI{3.2}{} & \SI{5}{}\\
			
			\rule{0pt}{10pt} S09 N1    & \SI{5}{} & \SI{20}{} & \SI{3700}{} & \SI{7.30}{} & \SI{0.23}{} & \SI{4.483}{} & \SI{16}{} & \SI{1.1}{} & \SI{2}{}\\
			\rule{0pt}{10pt} S09 N2    & \SI{5}{} & \SI{20}{} & \SI{3290}{} & \SI{7.30}{} & \SI{0.31}{} & \SI{4.973}{} & \SI{16}{} & \SI{1.4}{} & \SI{2}{}\\
			\rule{0pt}{10pt} S09 N3    & \SI{5}{} & \SI{20}{} & \SI{2950}{} & \SI{7.30}{} & \SI{1.18}{} & \SI{4.883}{} & \SI{16}{} & \SI{19.2}{} & \SI{30}{}\\
			\rule{0pt}{10pt} S09 N4    & \SI{5}{} & \SI{20}{} & \SI{2650}{} & \SI{7.30}{} & \SI{0.37}{} & \SI{6.046}{} & \SI{16}{} & \SI{1.3}{} & \SI{1}{}\\
			
			\rule{0pt}{10pt} S10 N1    & \SI{5}{} & \SI{20}{} & \SI{3700}{} & \SI{7.30}{} & \SI{0.27}{} & \SI{4.459}{} & \SI{19}{} & \SI{1.9}{} & \SI{4}{}\\
			\rule{0pt}{10pt} S10 N2    & \SI{5}{} & \SI{20}{} & \SI{3290}{} & \SI{7.30}{} & \SI{0.30}{} & \SI{4.978}{} & \SI{19}{} & \SI{2.2}{} & \SI{3}{}\\
			
			\rowcolor{lightgray} \rule{0pt}{10pt} S11 N2    & \SI{5}{} & \SI{20}{} & \SI{3430}{} & \SI{5.31}{} & \SI{0.28}{} & \SI{5.262}{} & \SI{20}{} & \SI{8.2}{} & \SI{10}{}\\ 
			\rule{0pt}{10pt} S11 N3    & \SI{5}{} & \SI{20}{} & \SI{3030}{} & \SI{5.31}{} & \SI{0.33}{} & \SI{5.873}{} & \SI{20}{} & \SI{14.9}{} & \SI{15}{}\\
			\rule{0pt}{10pt} S11 N4    & \SI{5}{} & \SI{20}{} & \SI{2713}{} & \SI{5.31}{} & \SI{0.22}{} & \SI{6.654}{} & \SI{20}{} & \SI{13.8}{} & \SI{11}{}\\ 
			
		\end{tabular}
	\end{adjustbox}
	\caption{Overview of nanowire resonator parameters. In this paper we focus on the nanowire resonator denoted S11 N2. The designed width $w$, spacing $s$ and length $l$ of the CPW resonator as well as the kinetic sheet inductance per square of the NbTiN film $L_\textrm{k}$, the extracted inductance of the nanowire at \SI{0}{\volt}, the measured resonance frequency at \SI{0}{\volt}, the voltage range $\Delta V_\textrm{g}$, the maximal change in frequency $\Delta f_\textrm{NW}$ and the maximal change in inductance $\Delta L_\textrm{NW}$ are given.}
	\label{tab: nanowire resonator parameters overview}
\end{table}

\begin{table}[]
	\centering
	\begin{tabular}{c|c|c|c|c|c}
		\rule{0pt}{10pt} label & $w$      & $s$      & $l$        & $L_\textrm{k}$              & $f_\textrm{r}$       \\ \hline
		\rule{0pt}{10pt}       & \si{\micro \meter}  & \si{\micro \meter}   & \si{\micro \meter}     & \si{\pico \henry \, \Box^{-1}}             & \si{\giga \hertz}         \\ \hline \hline
		
		\rule{0pt}{10pt} S04 R1    & \SI{5}{} & \SI{20}{} & \SI{2523}{} & \SI{5.45}{}    & \SI{7.423}{} \\
		\rule{0pt}{10pt} S04 R2    & \SI{5}{} & \SI{20}{} & \SI{2573}{} & \SI{5.55}{}    & \SI{7.239}{} \\
		
		\rule{0pt}{10pt} S06 R1    & \SI{5}{} & \SI{20}{} & \SI{2523}{} & \SI{5.57}{}    & \SI{7.372}{} \\
		\rule{0pt}{10pt} S06 R2    & \SI{5}{} & \SI{20}{} & \SI{2573}{} & \SI{5.77}{}    & \SI{7.155}{} \\
		
		\rule{0pt}{10pt} S07 R1    & \SI{4}{} & \SI{20}{} & \SI{4679}{} & \SI{5.45}{}    & \SI{3.798}{} \\
		
		\rule{0pt}{10pt} S09 R1    & \SI{5}{} & \SI{20}{} & \SI{4600}{} & \SI{7.3}{}     & \SI{3.704}{} \\
		\rule{0pt}{10pt} S09 R2    & \SI{5}{} & \SI{20}{} & \SI{2870}{} & \SI{7.4}{}     & \SI{5.910}{} \\
		
		\rule{0pt}{10pt} S10 R1    & \SI{5}{} & \SI{20}{} & \SI{4600}{} & \SI{7.3}{}     & \SI{3.709}{} \\
		\rule{0pt}{10pt} S10 R2    & \SI{5}{} & \SI{20}{} & \SI{2870}{} & \SI{7.4}{}     & \SI{5.918}{} \\
		
		\rowcolor{lightgray} \rule{0pt}{10pt} S11 R1    & \SI{5}{} & \SI{20}{} & \SI{4662}{} & \SI{5.35}{}    & \SI{4.042}{} \\ 
		\rule{0pt}{10pt} S11 R2    & \SI{5}{} & \SI{20}{} & \SI{4914}{} & \SI{5.33}{}    & \SI{3.838}{} \\
		\rule{0pt}{10pt} S11 R3    & \SI{5}{} & \SI{20}{} & \SI{2561}{} & \SI{5.34}{}    & \SI{7.360}{} \\
		\rule{0pt}{10pt} S11 R4    & \SI{5}{} & \SI{20}{} & \SI{2491}{} & \SI{5.22}{}    & \SI{7.616}{} \\
		
	\end{tabular}
	\caption{Overview of resonator parameters for all measured reference resonators labelled as sample SX with resonator RY. The designed width $w$, spacing $s$ and length $l$ of the CPW resonator as well as the extracted kinetic sheet inductance per square of the NbTiN film $L_\textrm{k}$, and the measured resonance frequency $f_\textrm{r}$ are given. In this paper we focus on the reference resonator denoted S11 R1.}
	\label{tab: reference resonator parameters overview}
\end{table}

In Fig. \ref{fig: statistics} we compare the gate voltage dependence of 12 nanowire resonators. We observe a monotonic increase in resonance frequency as a function of gate voltage. The frequency shift is quantified in Tab. \ref{tab: nanowire resonator parameters overview} as $\Delta f_\textrm{NW}$ over a voltage range of  $\Delta V_\textrm{g}$. The different magnitude of the frequency shift in the gate voltage scan arises from different participation ratios of nanowire to total inductance, but might also be a result of fluctuations between nanowires or due to the orientation of the Al shell relative to the bottom gate, potentially leading to gate screening for unfavourable orientations of the Al shell. We could not track the position of the Al shell relative to the bottom gate for each nanowire because of top gates encapsulating the nanowire. 

The 12 measured nanowires differ in their configuration: they comprise devices with different gate geometries, different nanowire growths, and measurement in different cooldowns. In particular, S04 N1 and S06 N2 use a \SI{2}{\micro \meter} long nanowire segment with a nanowire diameter of \SI{80(5)}{\nano \meter}, and a nominal thickness of the epitaxially grown Al of \SI{6}{\nano \meter}. 
However, they all exhibit similar gate responses. From this observation, we conclude that the effect of gate-tunable kinetic inductance is reproducible in hybrid superconducting semiconducting nanowires from Al and InAs and not unique to a specific nanowire batch or fabrication process.

Fig. \ref{fig: statistics_inductance} shows the decreasing nanowire inductance versus gate voltage for 12 nanowire resonators. The change in inductance is quantified in Tab. \ref{tab: nanowire resonator parameters overview} as $\Delta L_\textrm{NW}$. Overall, the inductance values are similar within a factor of 2 for most nanowires. The variations between the experiments probably arise from different residual inductances at the NbTiN to nanowire interface. These residual inductances do not appear in the reference resonators and thus are specific to the nanowire resonator.

\subsection{Reference resonators} \label{app: reference resonator}
\begin{figure}
	\includegraphics{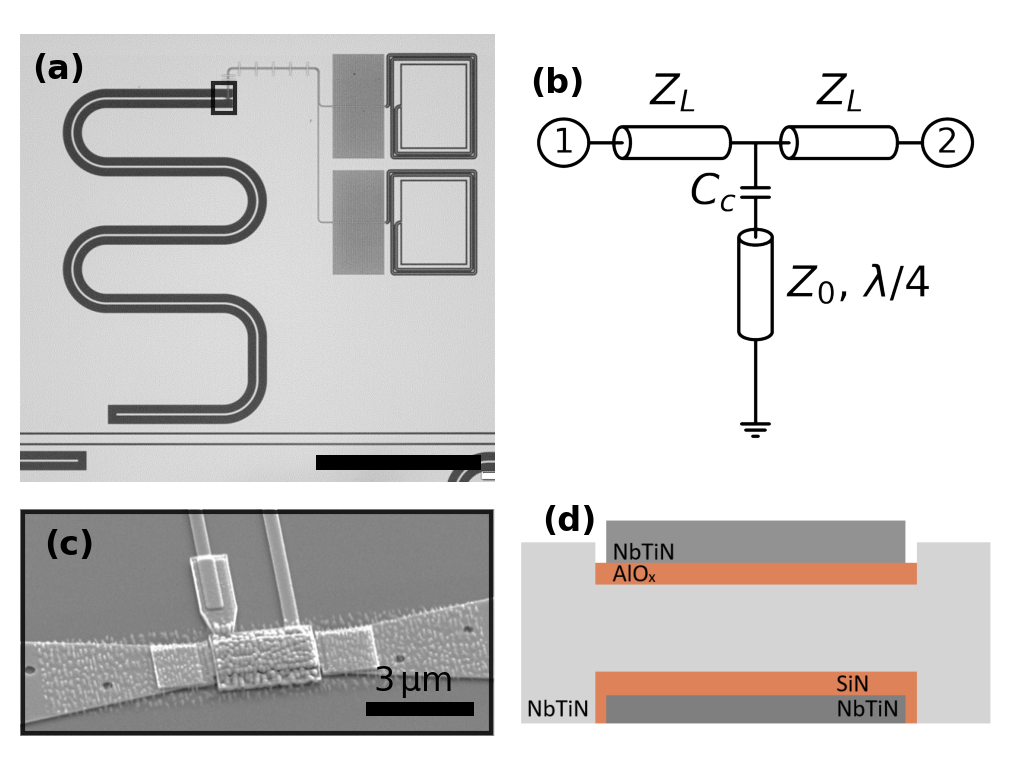}
	\caption{Measured reference device. (a) Optical image of one quarter-wave coplanar waveguide resonator capacitively coupled to a feedline and shorted to ground via a thick NbTiN film instead of a proximitized nanowire. Two LC filtered pads left to the resonator connect to the electrostatic gates in the vicinity of the NbTiN film. (scale bar \SI{300}{\micro \meter}) (b) Circuit diagram of a measured device. The transmission line resonator with impedance $Z_0$ is shunted to ground by a thick NbTiN film, and capacitively coupled ($C_\textrm{C}$) to the feedline with characteristic impedance $Z_\textrm{L}=\SI{50}{\ohm}$. Port 1 and 2 connect to the external measurement setup. 
		(c) Micrograph of the NbTiN film which is galvanically connected to the central conductor of the resonator [black box in (a)]. The \SI{3}{\micro \meter} long NbTiN patch is encapsulated by voltage bottom and top gates. Here, neither of the gates is used.
		(d) Schematic longitudinal cut away of reference structure used for comparison with the nanowire resonator. The nanowire section is replaced by a continuous \SI{150}{\nano \meter} thick NbTiN film. }
	\label{fig: reference resonator}
\end{figure}

The reference resonator consists of a quarter-wave coplanar waveguide resonators capacitively coupled to a feedline and shorted to ground via a thick NbTiN patch instead of a proximitized nanowire as shown in Fig. \ref{fig: reference resonator}a. Two LC filtered pads left to the resonator connect to the electrostatic gates in the vicinity of the NbTiN patch. However, these gates are not used.
The micrograph of the NbTiN film in Fig. \ref{fig: reference resonator}c shows the \SI{3}{\micro \meter} long and \SI{1}{\micro \meter} wide NbTiN patch encapsulated by voltage bottom and top gates to mimic the same electromagnetic environment as for the nanowire resonator.
The cut away in Fig. \ref{fig: reference resonator}d illustrates the difference to the nanowire resonator shown in the main text in Fig. 1d. This direct comparison allows us to test the device quality in this multi-step fabrication and the superconducting NbTiN to NbTiN contact.
The reference resonator can be modelled with the equivalent circuit in Fig. \ref{fig: reference resonator}b. The transmission line resonator with characteristic impedance $Z_0$ identical to the nanowire resonator, but different resonator impedance is capacitively coupled ($C_\textrm{C}$) to the feedline with characteristic impedance $Z_\textrm{L}=\SI{50}{\ohm}$. Port 1 and 2 connect to the external measurement setup. 

For the experiment, we design CPW resonators with identical transmission line impedance as the nanowire resonator, but different length $l$. The resonator parameters are listed in Tab \ref{tab: reference resonator parameters overview}. We extract the kinetic sheet inductance per square of the NbTiN film $L_\textrm{k}$ from the reference resonator by comparing the measured resonance frequencies with microwave simulations (AWR Microwave Office) of the full resonator design and numerical calculations based on a conformal mapping technique for CPW transmission lines \cite{Simons2001}. We obtain the kinetic sheet inductance from the average over all reference resonators per sample.

\subsection{Geometric inductance of nanowire segment} \label{app: geometric inductance}
We estimate the geometric inductance of the nanowire segment using a conformal mapping technique \cite{Simons2001}. For a \SI{100}{\nano \meter} wide and \SI{3}{\micro \meter} long wire in a CPW geometry with \SI{22}{\micro \meter} spacing to ground, we find a geometric inductance of \SI{4.5}{\pico \henry}. This is about \SI{2}{\percent} of the total nanowire inductance. This systematic error falls into the uncertainty of the nanowire inductance, thus we neglect the geometric contribution arising from the magnetic self-inductance.   

\subsection{Resonator bifurcation} \label{app: Resonator bifurcation}
\begin{figure}	
	\includegraphics{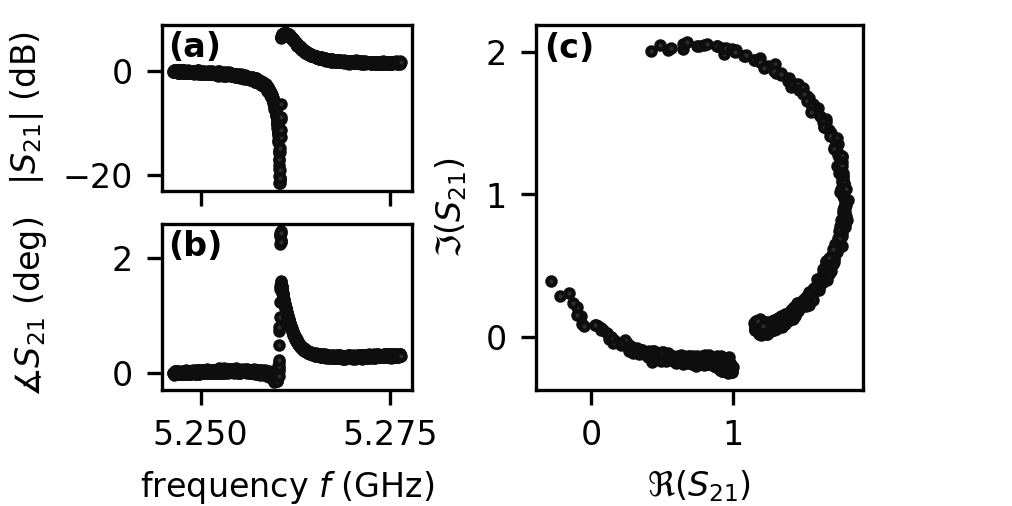}
	\caption{Scattering parameter $S_{21}$ of bifurcating nanowire resonance in magnitude (a), phase (b) versus frequency and in the complex plane (c) at \SI{-113}{dBm} input power at the reference plane of the resonator.}
	\label{fig: bifurcation}
\end{figure}
Superconducting resonators can exhibit nonlinear behaviour including bifurcation if driven with sufficiently large input power \cite{Swenson2013}. Fig. \ref{fig: bifurcation} shows the scattering parameter of a nanowire resonator in magnitude (a), phase (b) versus frequency and in the complex plane (c) at \SI{-113}{dBm} input power at the reference plane of the resonator. The discontinuity of the circle in Fig. \ref{fig: bifurcation} indicates a bifurcation behaviour at high readout powers. In the main text we limit our studies to the regime of few intra-cavity photons where the signal to noise ratio is sufficient for readout, but where the resonator does not bifurcate.

\subsection{Gate leakage} \label{app: gate leakage}
\begin{figure}
	\includegraphics{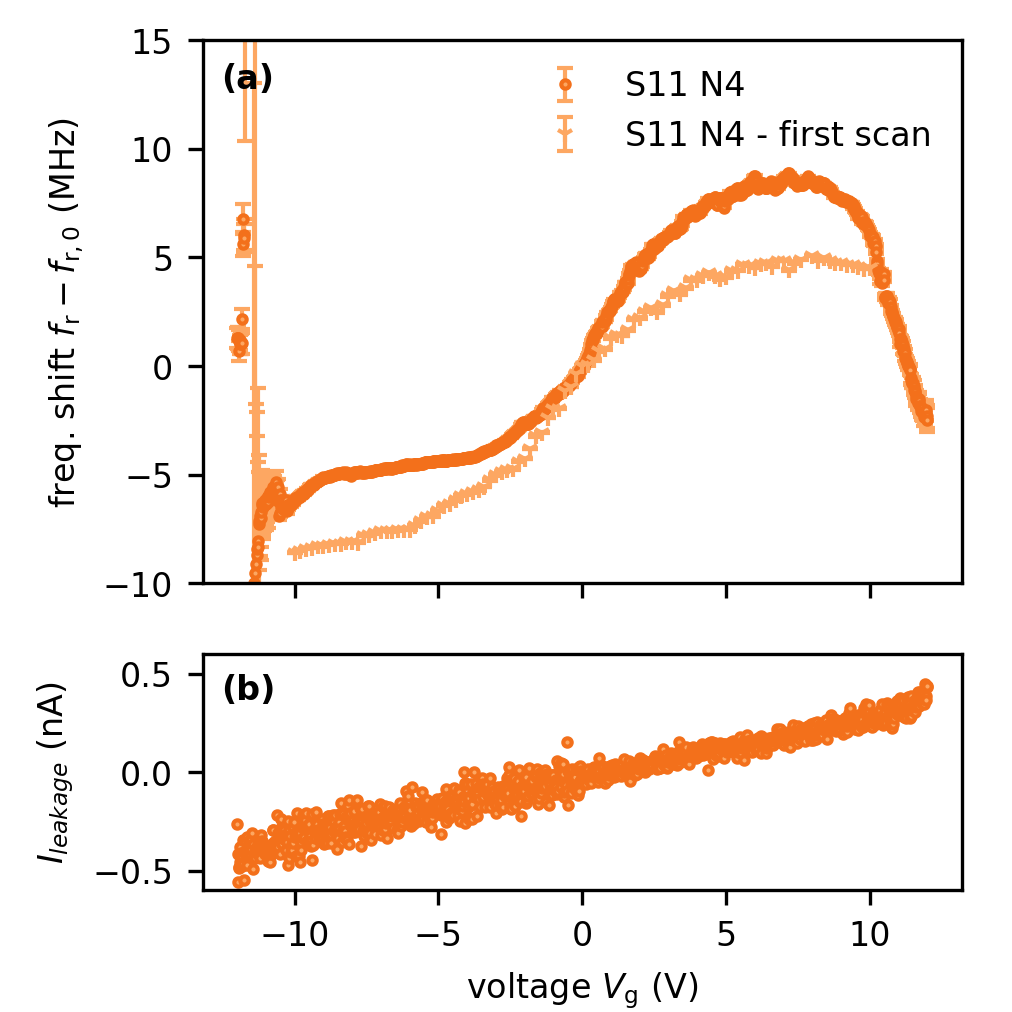}
	\caption{Gate leakage of nanowire resonator 4. (a) Frequency shift as function of gate voltages extracted from fits to $S_{21}$ data for resonator S11 N4 in the first scan of the experiment and later after exceeding a gate voltage of $|\SI{10}{\volt}|$. (b) Leakage current versus gate voltage measured during voltage sweep in the latter case.}
	\label{fig: gate leakage}
\end{figure}
Fig. \ref{fig: gate leakage}a shows the frequency shift of about \SI{13}{\mega \hertz} as function of gate voltages for nanowire resonator S11 N4 in the first scan of the experiment and after exceeding a gate voltage of $|\SI{10}{\volt}|$ multiple times. Between the two experiments, we observe a different slope around \SI{0}{\volt}, but otherwise the same magnitude of the frequency shift. 

In Fig. \ref{fig: gate leakage}b, while measuring the frequency response of S11 N4 between \SI{-12}{\volt} and \SI{+12}{\volt}, we record the leakage current through the bottom gate, which increases linearly by about \SI{1}{\nano \ampere}. 
We attribute the sharp drop in internal quality factor (see Fig. 3a) at large applied gate voltages to the point where the injection of quasi-particles through the bottom gate reduces the superconducting gap measurably. This is in line with the slightly asymmetric onset in the reduction of the internal quality factor for positive and negative voltages.  
The data shown in the main text has been taken after exceeding voltages of $|\SI{10}{\volt}|$. We note however that the trends in frequency and internal quality factor versus the control parameters do not seem to be affected in the voltage range from \SI{-8}{\volt} to \SI{8}{\volt}. 

\subsection{Gate stability and hysteresis} \label{app: stability}
\begin{figure}
	\includegraphics{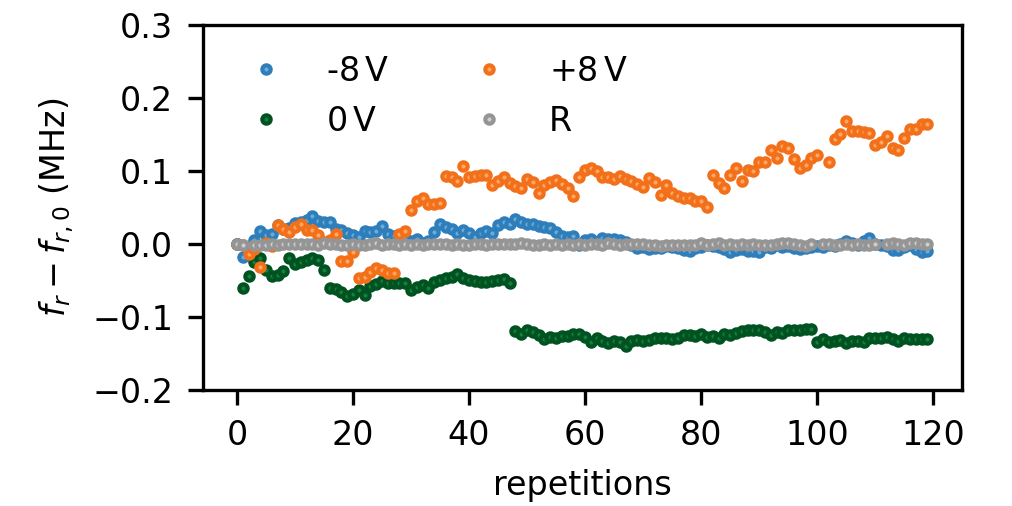}
	\caption{Gate voltage stability. Frequency shift of nanowire resonator for three different gate voltages and frequency shift of reference resonator for 120 repetitions with \SI{30}{s} waiting time between measurement point. }
	\label{fig: stability}
\end{figure}
\begin{figure}
	\includegraphics{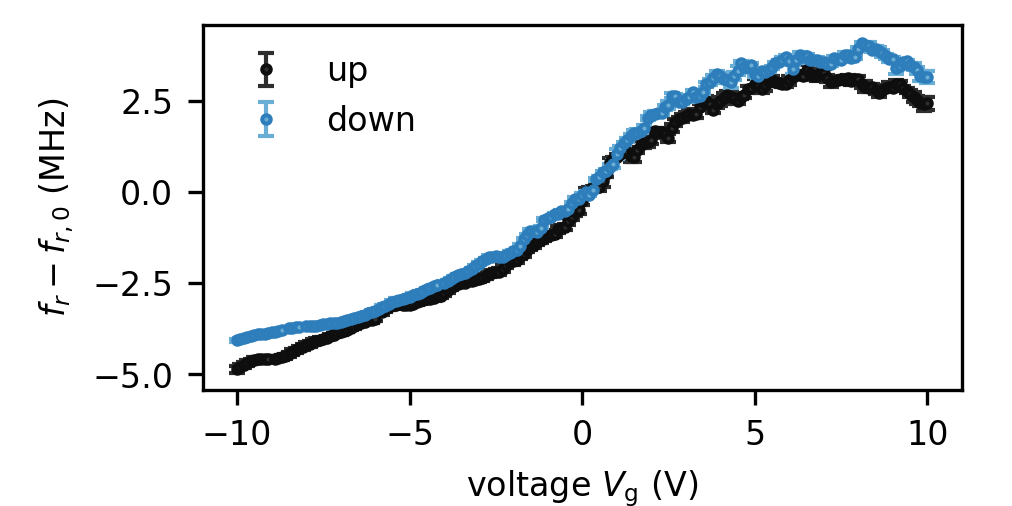}
	\caption{Gate voltage hysteresis. Average frequency shift of nanowire resonator over 4 gate voltage scans per sweep direction.}
	\label{fig: hysteresis}
\end{figure}
In view of possible applications for frequency tunable resonators we study the frequency stability over time and the gate hysteresis. Fig. \ref{fig: stability} shows the frequency shift of the nanowire resonator S11 N2 for three different gate voltages and the frequency shift of the reference resonator S11 R1 for 120 repetitions with \SI{30}{s} waiting time between measurements point. The reference resonator exhibits a constant frequency in time. Only small variations on the few kHz scale can be observed. The behaviour of the nanowire resonator however differs from the reference resonator. For negative gate voltages, the fluctuations in time are on the order of \SI{50}{\kilo \hertz}, but the trend is  relatively flat otherwise. The more positive we set the gate voltage, the more jumps and drifts appear in the time dependence. We speculate that at this gate voltage sudden reconfigurations of the charge carrier distribution in the nanowire happen. 

Fig. \ref{fig: hysteresis} shows the average frequency shift of the nanowire resonator for four gate voltage scans per sweep direction with a reset to zero volt between each scan. The error bar represents the standard deviation and is small compared to the total frequency shift. The size of the error bar is nearly equal over the entire range of gate voltages. We note the offset between the up and down sweeps originating from gate hysteresis or a reconfiguration of the charge carrier distribution.

\bibliography{apssamp} 

\end{document}